\documentclass[conference]{IEEEtran}
\makeatletter
\def\ps@headings{%
\def\@oddhead{\mbox{}\scriptsize\rightmark \hfil \thepage}%
\def\@evenhead{\scriptsize\thepage \hfil \leftmark\mbox{}}%
\def\@oddfoot{}%
\def\@evenfoot{}}
\makeatother
\ifCLASSINFOpdf
  \usepackage[pdftex]{graphicx}
\else
  \usepackage[dvips]{graphicx}
\fi

\usepackage[cmex10]{amsmath}
\usepackage{algorithmic}

\usepackage{array}

\usepackage[tight,footnotesize]{subfigure}
\usepackage[caption=false,font=footnotesize]{subfig}

\usepackage{url}
\usepackage{ifpdf}

\begin{document}

\title{\Large Throughput Optimal Flow Allocation on Multiple Paths \\ for Random Access Wireless Multi-hop Networks}

\author{\authorblockN{Manolis Ploumidis\textsuperscript{\dag}, Nikolaos Pappas\textsuperscript{\ddag}, Apostolos Traganitis\textsuperscript{\dag} \\}
\IEEEauthorblockA{\textsuperscript{\dag}Computer Science Department, University of Crete, Greece\\Institute of Computer Science, Foundation for Research and Technology - Hellas (FORTH)\\
\textsuperscript{\ddag} Sup\'{e}lec, Department of Telecommunications, Gif-sur-Yvette, France\\
\authorblockA{\textsuperscript{\dag}ploumid@ics.forth.gr,\textsuperscript{\ddag}nikolaos.pappas@supelec.fr,\textsuperscript{\dag}tragani@ics.forth.gr\\}}}


\maketitle

%********************************************************************************************************************************************************************* Abstract
\begin{abstract}
In this paper we consider random access wireless multi-hop mesh networks with multi-packet reception capabilities where multiple flows are forwarded to the gateways through node disjoint paths. 
We address the issue of aggregate throughput-optimal flow rate allocation with bounded delay guarantees.
We propose a distributed flow rate allocation scheme that formulates flow rate allocation as an optimization problem and derive the conditions for non-convexity for an illustrative topology.
We also employ a simple model for the average aggregate throughput achieved by all flows that captures both intra- and inter-path interference.
The proposed scheme is evaluated through NS-2 simulations.
Our preliminary results are derived from a grid topology and show that the proposed flow allocation scheme slightly underestimates
the average aggregate throughput observed in two simulated scenarios with two and three flows respectively.
Moreover it achieves significantly higher average aggregate throughput than single path utilization in two different traffic scenarios examined.
\end{abstract}

\IEEEpeerreviewmaketitle

%********************************************************************************************************************************************************************* Introduction
%\vspace{-0.2in}
\section{Introduction}

In order to better utilize the scarce resources of wireless multi-hop networks and meet the increased user demand for QoS, numerous studies have suggested the use of multiple paths in parallel.
Utilization of multiple paths can provide a wide range of benefits in terms of, throughput \cite{6133896}, delay \cite{mp_route_wim}, reliability \cite{mp_route_mpdsr}, load balancing \cite{5198989, 6133896}, security \cite{ref_mhrp} and energy efficiency \cite{5198989,5425265}.
However multipath utilization in wireless networks is more complicated compared to their wired counterparts since transmissions across a link interfere with neighboring links reducing thus network capacity.
As far as flow allocation on multiple paths and rate control is concerned, a well studied approach associates a utility function to each flow's rate and aims at maximizing the sum of these utilities subject to cross-layer constraints \cite{flow_alloc_2008, 4509706}.
Following \cite{kelly98ratecontrol}, the Network Utility Maximization (NUM) framework has found many applications in rate control over wireless networks \cite{4712692,conf/icc/QiuBX12}.
Authors in \cite{5089987}, instead of employing a utility function of a flow's rate, they employ a utility function of flow's effective rate in order to take into account the effect of lossy links along a multi-hop.
Based on the theoretical ideas of back-pressure scheduling and utility maximization Horizon \cite{horizon_2008} constitutes a practical implementation of a multipath forwarding scheme that interacts with TCP.

In this study we address the issue of aggregate throughput-optimal flow rate allocation for random access wireless multi-hop mesh networks with multi-packet reception capabilities where multiple flows are forwarded to the gateways through node disjoint paths.
The main contribution of this study is a distributed  flow rate allocation scheme that formulates flow rate allocation as an optimization problem.
The key feature of the proposed scheme is that it is aimed at maximizing the average aggregate throughput achieved by all flows as opposed to most the of the aforementioned studies that consider a sum of utility functions assigned to each flow's rate.
Moreover, the suggested scheme also guarantees bounded packet delay.
The second contribution of this study is a simple model for the average aggregate throughput, capturing both inter- and intra-path interference through the SINR model.

The suggested scheme is demonstrated through a set of toy topologies.
We present the conditions under which the corresponding optimization problem is non-convex for one toy topology.
We evaluate the suggested flow allocation scheme through NS-2 simulations.
Our preliminary results reveal that our scheme slightly under-estimates the actual average aggregate throughput (AAT)
by 5.1\% and 3.7\% on average for two simulated scenarios with two and three flows respectively.
The reason for this under-estimation is explained in section \ref{sec:simulation_results}.
Moreover, we compare the AAT achieved by the proposed scheme with a single-path flow allocation scheme that optimally utilizes the best path available to the destination.
Our scheme achieves 49.1\% and 102.8\% higher AAT on average for the cases of two and three flows respectively.

The rest of the paper is organized as follows: Section \ref{sec:system_model} presents the system model considered, while section \ref{sec:analysis}
presents how aggregate throughput optimal flow rate allocation is formulated as an optimization problem. In section \ref{sec:simulation_results}
we present some preliminary results and conclude in \ref{sec:conclusions}.

%%********************************************************************************************************************************************************************* System Model
%%\vspace{-0.1in}
\section{System Model}
\label{sec:system_model}

We consider static wireless multi-hop networks with the following properties:
\begin{IEEEitemize}
    \item Random access to the shared medium where each node transmits independently of all other nodes requiring no coordination among them.
    When relay nodes are concerned, $q_{i}$ is used to denote the packet transmission probability for node \textit{i} given there is a packet available for transmission.
    For flow originators it denotes the rate at which they inject packets into the network (flow rate).
    For relay nodes, $q_i$ is assumed to be a parameter whose value is fixed and propagated periodically back to the sources through routing protocol's control messages.
    \item Time is slotted and slot length equals the time required to transmit a single packet.
    \item Flows carry unicast traffic of same size packets.
    \item All nodes use the same channel and rate, and are equipped with multi-user detectors being thus able to successfully decode packets
    from more than one transmitter at the same slot \cite{Verdu:1998:MD:521411}.
    \item Each node cannot send and receive simultaneously.
    \item All nodes always have packets available for transmission.
    \item Propagation delay among nodes and queueing delay are ignored.
    \item For each node its position, transmission probability or flow rate along with an indication of whether it is a flow originator are assumed known to all other nodes.
    This information can be periodically propagated throughout the network through a link-state routing protocol.
    \item Concerning the routing protocol, we assume that it performs source routing ensuring that packets of the same
    flow will be forwarded to the destination through the same path, and that it provides to traffic sources multiple node- and link-
    disjoint paths (referred to as \textit{disjoint} paths for the rest of the paper).
\end{IEEEitemize}

\subsection{Physical Layer Model}

The MPR channel model used in this paper is a generalized form of the packet erasure model.
In the wireless environment, a packet can be decoded correctly by the receiver if the received $SINR$ exceeds a certain threshold.
More precisely, suppose that we are given a set $T$ of nodes transmitting in the same time slot.
Let  $P_{rx}(i,j)$ be the signal power received from node $i$ at node $j$.
Let $SINR(i,j)$ be expressed using (\ref{eq:sinr_thres}).
\begin{equation}
\label{eq:sinr_thres}
SINR(i,j)=\frac{P_{rx}(i,j)}{\eta_{j}+\sum_{k\in T\backslash\left\{i\right\}} {P_{rx}(k,j)}}
\end{equation}
In the above equation $\eta_{j}$ denotes the receiver noise power at $j$. We assume that a packet transmitted by $i$ is successfully received by $j$
if and only if $SINR(i,j)\geq \gamma_{j}$, where $\gamma_{j}$ is a threshold characteristic of node $j$. The wireless channel is subject to fading;
let $P_{tx}(i)$ be the transmitting power of node $i$ and $r(i,j)$ be the distance between $i$ and $j$. The power received by $j$ when $i$
transmits is $P_{rx}(i,j)=A(i,j)g(i,j)$ where $A(i,j)$ is a random variable representing channel fading. Under Rayleigh fading, it is
known~\cite{b:Tse} that $A(i,j)$ is exponentially distributed. The received power factor $g(i,j)$ is given by $g(i,j)=P_{tx}(i)(r(i,j))^{-\alpha}$
where $\alpha$ is the path loss exponent with typical values between $2$ and $4$. The success probability of link $(i,j)$ when the transmitting nodes
are in $T$ is given by

{\scriptsize
\begin{equation}
\label{eq:succprob}
p_{i/T}^{j}=\exp\left(-\frac{\gamma_{j}\eta_{j}}{v(i,j)g(i,j)}\right) \prod_{k\in T\backslash \left\{i,j\right\}}{\left(1+\gamma_{j}\frac{v(k,j)g(k,j)}{v(i,j)g(i,j)}\right)}^{-1}
\end{equation}
}
where $v(i,j)$ is the parameter of the Rayleigh random variable for fading. The analytical derivation for this success probability
which captures the effect of interference on link $(i,j)$ from transmissions of nodes in set $T$, can be found in~\cite{b:Nguyen}.

%********************************************************************************************************************************************************************* Analysis
\section{Analysis}
\label{sec:analysis}

In this section we present how aggregate throughput optimal flow rate allocation is formulated as an optimization problem for random topologies.
We also provide an illustrative example of the suggested scheme through a toy topology.
\subsection{Throughput optimal flow rate allocation}
\label{sec:analysis_generalized}

The suggested method for formulating aggregate throughput optimal flow rate allocation as an optimization problem for random topologies is a procedure consisting of three steps.
First, some notations should be introduced.
\textit{V} denotes the set of N network nodes.
We assume $m$ flows $f_{1}, f_{2},...,f_{m},$ that need to forward traffic to destination node $D$.
$R = \lbrace r_{1}, r_{2}, ...,r_{m} \rbrace$ represents the set of $m$ disjoint paths employed by these flows.
$|r_{i}|$ is used to denote the number of links in path $r_{i}$.
$I_{i,j}$ is the set of nodes that cause interference to packets sent from 'i' to 'j'.
For example, if the maximum number of interfering nodes is assumed and $j \neq D$, then $I_{i,j} = V \setminus \lbrace i, j, D \rbrace$ and thus link's (i,j) set of interfering nodes has size $L_{i,j}=|I_{i,j}| = |V|-3$.
Further on, $Src(r_{k})$ is used to denote the flow originator that employs path $r_{k}$.
$\bar{T}_{i,j}$ and $\bar{T}_{r_{k}}$ denote the average throughput measured in packets per slot achieved by link (i,j) and flow $f_{k}$ forwarded over path $r_{k}$ respectively.

The first step of the suggested method consists of deriving the expression for the average throughput of a random link (i,j).
Average throughput for that link, $\bar{T}_{i,j}$, can be expressed through (\ref{eq:process_step_1}).
\begin{equation}
\label{eq:process_step_1}
\begin{aligned}
\bar{T}_{i,j}  =  \sum_{ l=0 }^{ 2^{ L_{i,j} }-1 } P_{i,j,l}  q_{i,j}  \prod_{n=1}^{ L_{i,j} } q_{I_{i,j}[n]}^{b(l,n)}  (1 - q_{I_{i,j}[n]} )^{1-b(l,n)}\\
\end{aligned}
\end{equation}
where,
\begin{equation*}
\begin{aligned}
& q_{i,j}=\left\{\begin{matrix}
& q_{i} \quad & j =  D\\
& q_{i}(1-q_{j}) \quad & j\neq D\\
\end{matrix}\right.,\\
\end{aligned}
\end{equation*}
\begin{equation*}
\begin{aligned}
& P_{i,j,l}=p_{i/i\cup \{ I_{i,j}[n], \; \forall \; n: \; b(l,n) \neq0) \} }^{j}, \\
& b(l,n) = l \; \& \; 2^{n-1}, \; \text{\& is the logical bitwise AND operator.} \\
\end{aligned}
\end{equation*}

The outcome of a transmission along link (i,j) during a slot depends on the amount of interference experienced with interference
depending on the set of other transmitters that are active during the same slot.
A node 'i' is active during a slot with probability $q_{i}$.
For flow originators $q_{i}$ denotes flow rate.
As a flow's data rate is increased, the interference imposed on other links is also increased.
Estimating thus a link's (i,j) throughput requires enumerating all possible subsets of active transmitters.
Assuming the maximum number of interfering nodes and a network with N nodes, all such subsets of interfering nodes for (i,j) are $2^{L_{i,j}}$.
For large networks enumerating all subsets of active transmitters may be computationally intractable.
More efficient approaches however could take into account the subset of nodes that have the most significant contribution in terms of interference at the cost of lower accuracy.
In (\ref{eq:process_step_1}),  $l$ enumerates all possible subsets of active transmitters while $b(l,n)$ becomes one if the $n^{th}$ node in $I_{i,j}$ is assumed
active in the $l^{th}$ subset examined.

The average aggregate throughput achieved by all flows is expressed through $\bar{T}_{aggr} = \sum_{ k= 1 }^{m} \bar{T}_{r_{k} }$ where $\bar{T}_{r_{k}} = \underset{ (i,j) \in r_{k} }{ min }  \bar{T}_{i,j} $.
The second step of the suggested method consists of maximizing the average aggregate throughput while also guaranteeing bounded packet delay which results in non-smooth optimization problem P1:
\begin{equation*}
\begin{aligned}
\underset{S}{\text{Maximize}} \quad & \sum_{k=1}^{m} \underset{ (i,j) \in r_{k} }{ min } \bar{T}_{i,j} \quad \quad \quad \quad \quad \quad (P1)\\
\text{s.t:}  \quad & (S1):\;  0 \leq q_{Src(r_{k})} \leq 1, \; k=1...m\\
& (S2):\; \bar{T}_{i,j} \leq \bar{T}_{j,l}, \\
& \quad \quad \quad \lbrace \forall i,j,l,k :  (i,j), \; (j,l) \in r_{k}, |r_{k}| > 1 \rbrace \\
\end{aligned}
\end{equation*}
,where, $S=\lbrace q_{Src(r_{k})}, \; k=1...m \rbrace$.
Constraint set S1 ensures that the maximum data rate for any flow does not exceed one packet per slot while also allowing paths that are not optimal to use, to remain unutilized.
Constraint set S2 ensures that for all links along paths that consist of more than one hop outgoing flow is larger or equal to incoming, preventing the accumulation of packets at a relay node, which guarantees bounded packet delay.

P1 can be transformed to the following smooth optimization problem:
\begin{equation*}
\begin{aligned}
& \underset{S'}{\text{Maximize}} \sum_{k=1}^{m}\left\{\begin{matrix}
\bar{T}_{Src(r_{k}),D}, & |r_{k}|=1 \quad \quad \quad \quad \quad (P2)\\
q'_{Src(r_{k})}, & |r_{k}|>1 \quad \quad \quad \quad \quad \quad \quad \quad
\end{matrix}\right.\\
& s.t.: \\
& \quad \quad (S1): \; 0 \leq q_{Src(r_{k})} \leq 1, \; k=1...m\\
& \quad \quad(S2): \; \bar{T}_{i,j} \leq \bar{T}_{j,l}, \; \\
& \quad \quad \quad \quad \quad \lbrace \forall i,j,l,k :  (i,j), \; (j,l) \in r_{k}, |r_{k}| > 1 \rbrace \\
& \quad \quad(S3): \; 0 \leq  q'_{Src(r_{k}) } \leq 1, \; \lbrace \forall k :  |r_{k}|>1 \rbrace \\
& \quad \quad(S4): \; q'_{Src(r_{k}) } \leq \bar{T}_{i,j}, \; \lbrace \forall i,j,k :  |r_{k}|>1, \; (i,j) \in r_{k} \rbrace \\
\end{aligned} \\
\end{equation*}
where, $S'=\lbrace q_{Src (r_{k} )}, \;  k=1...m \rbrace \cup \lbrace q'_{src (r_{k} )} : \;  |r_{k}|>1 \rbrace$

Network settings where the aforementioned analysis and flow rate allocation scheme can be applied may be the following:
\begin{IEEEitemize}
    \item Nodes within the network generate flows whose aggregate throughput to network gateways needs to be maximized.
    Maximizing however the aggregate throughput achieved by all flows can result in unfair data rate allocation.
    Provision for fairness issues though is part of our future work.
    \item Flows forwarded through the mesh network are split to multiple flows in order to increase the throughput achieved
    \item In sensor networks where specific nodes collect and forward data on behalf of other nodes aiming at maximizing throughput at the sink
\end{IEEEitemize}
\subsection{Throughput optimal flow rate allocation: Illustrative scenario}
\label{sec:mpath_mhop_analysis}

\begin{figure}
\centering
\includegraphics[width=5.2cm, height=2.5cm]{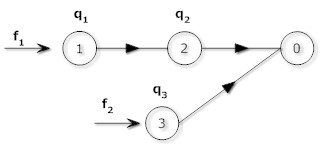}
\caption{Illustrative topology.}
\label{fig:mpath_mhop}
\end{figure}

We consider the toy topology presented in Fig. \ref{fig:mpath_mhop}.
Two flows namely, $f_{1}$ and $f_{2}$, originating at nodes 1, and 3 are forwarded to destination node 0 through paths $r_{1}$: 1-2-0 and $r_{2}$: 3-0 respectively.
We further assume that transmissions on a specific link cause interference to all other links.
Before presenting each link's average throughput consider link (2,0) as an example.
Transmitters that cause interference to packets sent from 2 to 0 constitute set $I_{2,0} = \lbrace 1,3 \rbrace$ and thus $L_{i,j}=2$.
There are four possible subsets of nodes that may cause interference on link (2,0) : $\lbrace \o  \rbrace, \lbrace 1  \rbrace , \lbrace 3  \rbrace ,\lbrace 1,3  \rbrace $.
When $l=3$ in (\ref{eq:process_step_1}), it enumerates the fourth subset of interfering nodes with $b(l,n)$ becoming one for both $n=1$ and $n=2$.

The average throughput per link is presented in (\ref{eq:multi_path_one_relay_linkthr}a)-(\ref{eq:multi_path_one_relay_linkthr}c).
\begin{subequations}
\label{eq:multi_path_one_relay_linkthr}
\begin{align}
{\bar{T}}_{1,2} &= q_{1}(1-q_{2})(1-q_{3})p_{1/1}^{2} + q_{1}(1-q_{2})q_{3}p_{1/1,3}^{2} \\
{\bar{T}}_{2,0} &= q_{2} (1-q_{1}) (1-q_{3}) p_{2/2}^{0} + q_{2} q_{1} (1-q_{3}) p_{2/2,1}^{0} \nonumber \\ &+ q_{2} (1-q_{1}) q_{3} p_{2/2,3}^{0} + q_{2} q_{1} q_{3} p_{2/1,2,3}^{0}\\
{\bar{T}}_{3,0} &= q_{3} (1-q_{1}) (1-q_{2}) p_{3/3}^{0} + q_{3} q_{1} (1-q_{2}) p_{3/1,3}^{0} \nonumber \\ &+ q_{3} (1-q_{1}) q_{2} p_{3/2,3}^{0} + q_{3} q_{1} q_{2}  p_{3/1,2,3}^{0}
\end{align}
\end{subequations}

Recall that $q_{1}$ and $q_{3}$ denote the data rates for flows $f_{1}$ and $f_{2}$ respectively.
Aggregate average throughput achieved by all flows can be expressed through (\ref{eq:one_relay_aggr_throughput}).
\begin{equation}
\label{eq:one_relay_aggr_throughput}
\begin{aligned}
& {\bar{T}}_{aggr} ={\bar{T}}_{r_{1}}+{\bar{T}}_{r_{2}}, \quad where, \\
& {\bar{T}}_{r_{1}} = min\{ {\bar{T}}_{1,2},{\bar{T}}_{2,0} \}, \quad {\bar{T}}_{r_{2}} = {\bar{T}}_{3,0}
\end{aligned}
\end{equation}

Aggregate throughput-optimal flow rate allocation consists of identifying rates $q_{1}$, $q_{3}$ that
maximize average aggregate throughput while also guaranteeing bounded packet delay.
These rates can be found by solving the following optimization problem: 
\begin{equation*}
\begin{aligned}
& \underset{q_{1},q_{3}}{\text{Maximize}}
& & {\bar{T}}_{30}+min\{ {\bar{T}}_{12},{\bar{T}}_{20} \}\\
& \text{subject to}
& & 0 \leq q_{i} \leq 1, \; i \in \lbrace 1,3 \rbrace \quad (g1)-(g4)\quad \quad \quad \\%(P3)\\
& & & {\bar{T}}_{12}\leq {\bar{T}}_{20} \quad \quad \quad \quad\quad \quad(g5)\\
\end{aligned}
\end{equation*}

Constraint (g5) constitutes the bounded delay constraint for path $r_{1}$.
According to third step of the process presented in the previous subsection, the above non-smooth optimization problem can be transformed to the following smooth optimization problem:
\begin{equation*}
\begin{aligned}
& \underset{q_{1}^{'},q_{1},q_{3} }{\text{Maximize}}
& & {\bar{T}}_{30}+q_{1}^{'} \\
& \text{subject to}
& & 0 \leq q_{i}\leq 1, \; i \in \lbrace 1,3 \rbrace \quad & (g1) &-(g4)\\
& & & {\bar{T}}_{12}\leq {\bar{T}}_{20}, & (g5)\\
& & & q_{1}^{'} \leq {\bar{T}}_{12}, & (g6) & \quad \quad \quad \quad (P3)\\
& & & q_{1}^{'} \leq {\bar{T}}_{20}, & (g7)\\
& & & 0 \leq q_{1}^{'} \leq 1 & (g8)&-(g9)\\
\end{aligned}
\end{equation*}

Optimization problem P3 is non-convex if the following condition holds: $p_{2/2,3}^{0} - p_{2/1,2,3}^{0} < \frac{ 1-q_{2} }{ q_{2} } (p_{1/1}^{0} - p_{1/1,3}^{2}) + p_{2/2}^{0} - p_{2/1,2}^{0}$.

Before presenting simulation results for a larger topology, we further motivate flow rate allocation on multiple paths using numerical results derived from the toy topology depicted in Fig~\ref{fig:mpath_mhop}.
Let $d(i,j)$ denote the distance between nodes $i$ and $j$.
Let also $P_{ r_{k} } = \prod_{ (i,j) \in r_{k} } p_{i/i}^{j}$ denote the \textit{end-to-end success probability} for path $r_{k}$.
For the illustrative purpose of this section we assume that $d(1,2)=d(2,0)=d(3,1)=d$, $d(3,0)=\sqrt 5d$, $d(3,2) = \sqrt 2 d$, where $d=400m$.
Further on, the path loss exponent assumed is $3$ while transmission probability for relay node $2$ is $0.5$.
Flow rates $q_{1}$ and $q_{3}$ that achieve maximum average aggregate throughput (AAT) for SINR threshold values $\lbrace 0.25, 0.5,...,2 \rbrace$ are estimated by solving optimization problem (P3) using the simulated annealing.
It should be noted that multi-hop path $r_{1}$ : 1-2-0 exhibits higher end-to-end success probability than path $r_{2}$ : 3-0 for all aforementioned SINR threshold values.
\begin{figure}
\centering
\includegraphics[width=9.2cm, height=6.5cm]{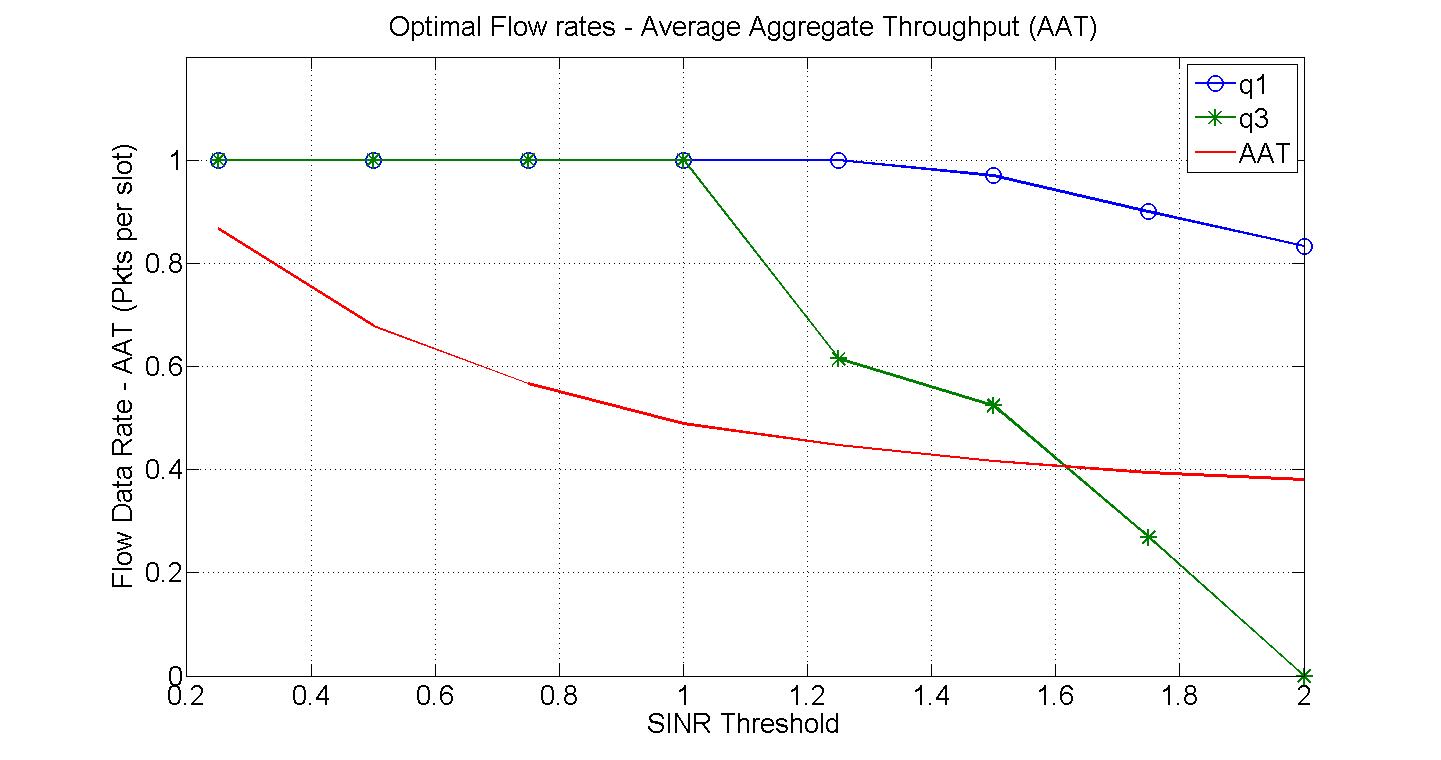}
\caption{Optimal Flow rates and Average Aggregate Throughput achieved.}
\label{fig:num_res_toy_topo}
\end{figure}

In Fig. \ref{fig:num_res_toy_topo} we present throughput optimal flow rates assigned on paths $r_{1}$, and $r_{2}$ along with the average aggregate throughput achieved (AAT) for the SINR threshold values considered.
As this figure shows, the maximum AAT is achieved by full rate utilization of both paths for SINR threshold values up to $1.0$ suggesting that inter-flow interference is balanced by the gain in throughput.
For SINR threshold values larger than $1.0$, utilization of path $r_{2}$ which exhibits lower performance in terms of end-to-end success probability declines.
This is due to the fact that for large SINR threshold values the effect of interference imposed on path $r_{1}$ becomes more significant.
At the same time, flow forwarded through path $r_{2}$ manages to deliver only a small portion of its traffic to destination node $0$.
%\vspace{-0.2in}
\section{Simulation Results}
\label{sec:simulation_results}

We evaluate the proposed aggregate throughput-optimal flow rate allocation scheme  using network simulator NS-2, version 2.34 \cite{ref:ns2}, including support for multiple transmission rates \cite{ref:dei80211mr}.
Following the assumptions of the system model, it employs a custom source-routed link-state routing protocol based on UM-OLSR \cite{ref:um-olsr}.
Topology control messages propagate each node i's position, transmission probability $q_{i}$ (or flow rate) and an indication of whether it is a flow originator throughout the network every 5 seconds.
Traffic sources employ static predefined routes to the destination and generate constant bit rate UDP flows.
Implementing a search algorithm for node-disjoint paths is out of the scope of the evaluation process.
Concerning medium access control, a slotted aloha-based MAC layer is implemented.
Transmission of data, routing protocol control and ARP packets is performed at the beginning of each slot without performing carrier sensing prior to transmitting.
Acknowledgments for data packets are sent immediately after successful packet reception while failed frames are retransmitted.
Slot length $T_{slot}$ is expressed through: $T_{slot} = T_{data} + T_{ack} + 2D_{prop}$ where $T_{data}$ and $T_{ack}$ denote the transmission times for data packets and ACKs while $D_{prop}$ denotes the propagation delay.
As far as physical layer is concerned, all packets including ACKs are successfully decoded if their received SINR exceeds the SINR threshold.
The received SINR for each packet is estimated through (\ref{eq:sinr_thres}) and path loss exponent is assumed to be $4$.
Transmitters during each slot that are considered to contribute with interference are those transmitting data packets or routing protocol control packets.
All nodes use the same SINR threshold, transmission rate and channel.
Transmission power and noise is 0.1 and $7 \cdot 10^{-11}$ Watt respectively.
For each relay node $i$ the transmission probability, given there is a packet for transmission, ($q_{i}$) is fixed to 0.5 for the whole simulation period.
Using the information propagated through topology control messages each node applies the method presented in Section \ref{sec:analysis_generalized} in order to formulate an instance of optimization problem P2 based on the topology infered.
Throughput optimal flow rates $q_{j}$ are then estimated on a distributed manner where each flow originator $j$ solves its own instance of optimization problem P2 using simulated annealing.

\begin{figure}
\centering
\includegraphics[width=6.0cm, height=4.5cm]{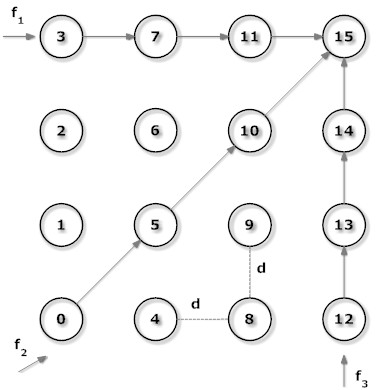}
\caption{Grid network topology.}
\label{fig:grid_topology}
\end{figure}

Our preliminary results are derived from the topology presented in Fig. \ref{fig:grid_topology} which is a grid consisting of 16 nodes.
Both vertical and horizontal distance for all pairs of nodes $d$ is set to 100meters.
The purpose of the evaluation process is two-fold.
Firstly we evaluate the accuracy of the average aggregate throughput (AAT) estimation by the suggested flow rate allocation scheme by comparing it with the corresponding throughput value obtained from simulated scenarios.
Secondly as a first step towards evaluating the effectiveness of the suggested scheme, we compare its performance in terms of ATT with \textit{Best-path} scheme.
Best-path utilizes a single path to forward data to the destination.
The data rate of the flow routed over that path is estimated by solving a single flow version of P2.
The path $p$ employed by Best-path is the one with the highest end-to-end success probability presented in section \ref{sec:mpath_mhop_analysis}.

\begin{figure}
\centering
\includegraphics[width=9.2cm, height=6.2cm]{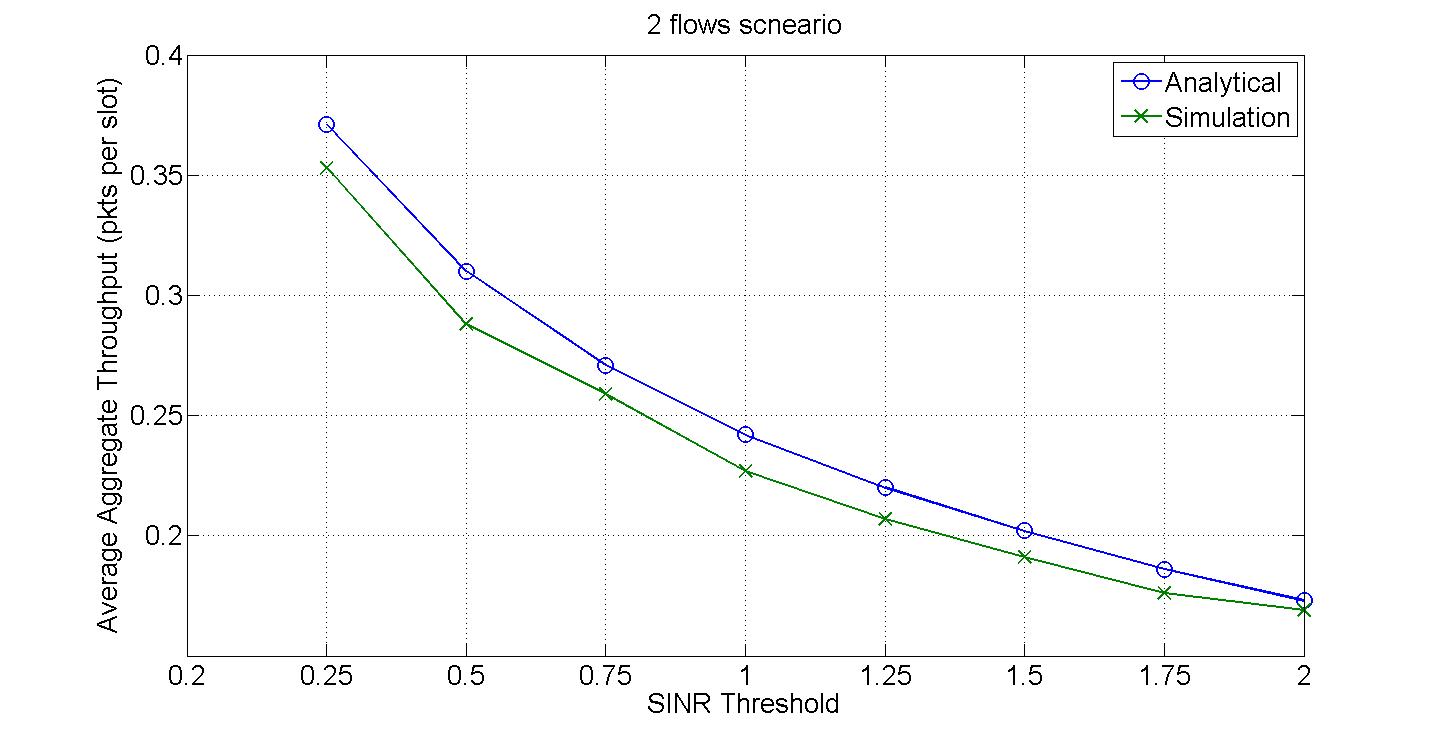}
\caption{Average Aggregate Throughput: Analytical vs. Simulation - Two flows scenario.}
\label{fig:fig_1}
\end{figure}

\begin{figure}
\centering
\includegraphics[width=9.2cm, height=6.2cm]{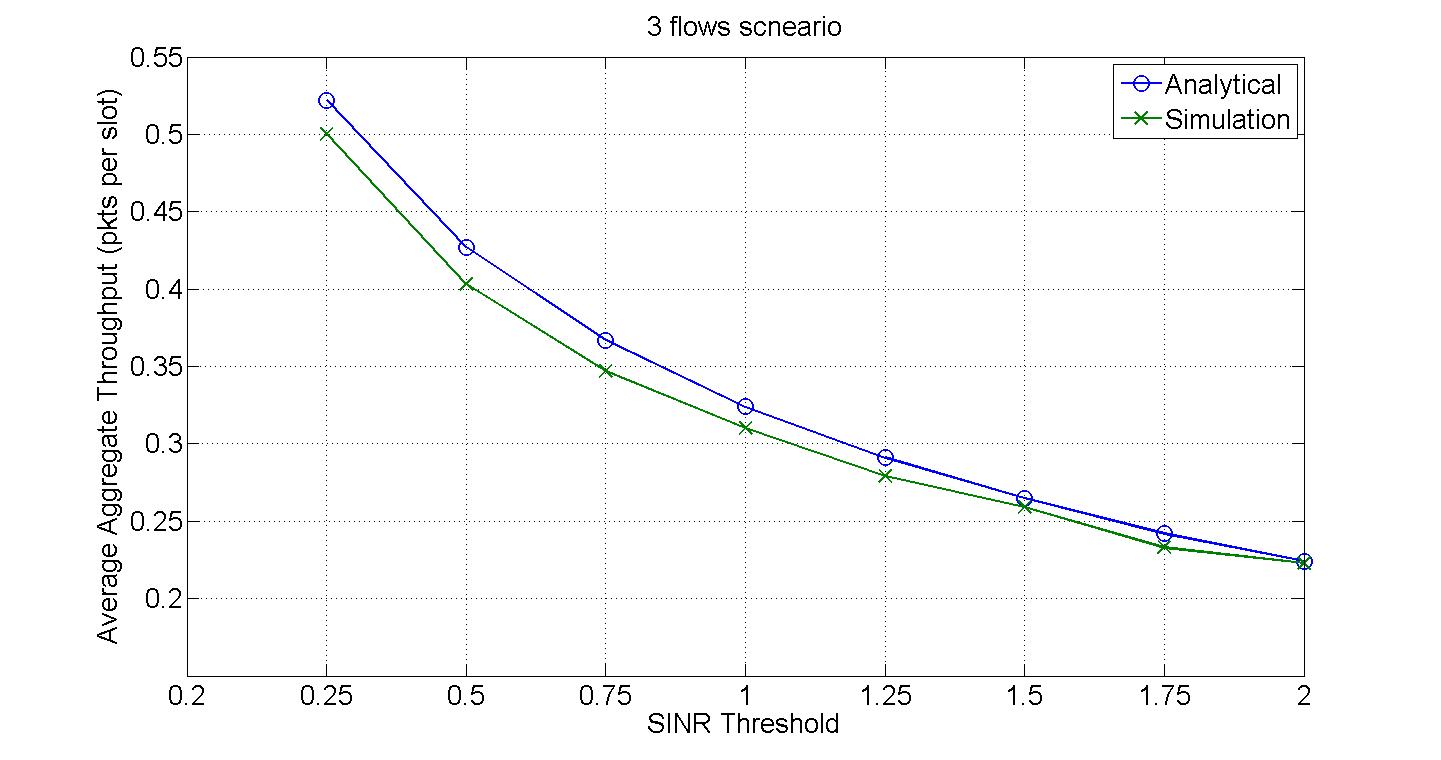}
\caption{Average Aggregate Throughput: Analytical vs. Simulation - Three flows scenario.}
\label{fig:fig_2}
\end{figure}

Two traffic scenarios are explored, also depicted in Fig. \ref{fig:grid_topology}.
In the first scenario, two flows $f_{1}$ and $f_{2}$, originated at nodes 3 and 0 respectively, are routed to destination node $15$ through paths $r_{1}$: 3-7-11-15 and $r_{2}$: 0-5-10-15.
In the second scenario, a third flow $f_{3}$ is also considered and it is routed through path $r_{3}$: 12-13-14-15.
The effect of interference on success probability and thus on throughput is captured by considering different values of the SINR threshold.
For each traffic scenario and SINR threshold value we derive a simulation scenario of 300 seconds where our flow rate allocation scheme is compared with Best-Path one.
Fig.~\ref{fig:fig_1} and~\ref{fig:fig_2} compare analytical with simulation results concerning average aggregate throughput for SINR threshold values $\lbrace 0.25, 0.5,...,2.0 \rbrace$.

As these figures show, the proposed aggregate throughput optimal flow data rate allocation scheme (also referred to as \textit{TODRA}) slightly underestimates average aggregate throughput.
More precisely, our scheme underestimates the actual throughput observed in simulated scenarios by 5.1\% and 3.7\% on overage concerning all SINR threshold values for the two and three flows scenarios.

There are three reasons for this underestimation.
First of all, in the system model and our analysis we have disregarded the acknowledgments sent by receivers upon successful packet reception and consequently their probability of being dropped due to low SINR.
In the scenarios simulated, it was observed that the lower the SINR threshold, the larger the number of successfully received packets and thus the larger the number of ACKs sent back to the transmitters.
This phenomenon results in ACKs being synchronized and finally dropped due to low received SINR.
The second reason for the aforementioned underestimation is that in our analysis we silently disregard control traffic due to routing protocol and ARP protocol and consider that all slots carry data packets.
Finally, in our analysis, we consider interference caused only by nodes participating in paths employed by the flows forwarded.
However, network nodes not participating in the multipath employed periodically transmit control traffic and thus contribute to interference.
The two last reasons for throughput underestimation are less important however due to the large intervals over which control packets are generated.
Consequently, the simplifying assumptions adopted in our analysis  cause an insignificant underestimation of the average aggregate throughput.
\begin{figure}
\centering
\includegraphics[width=9.2cm, height=6.2cm]{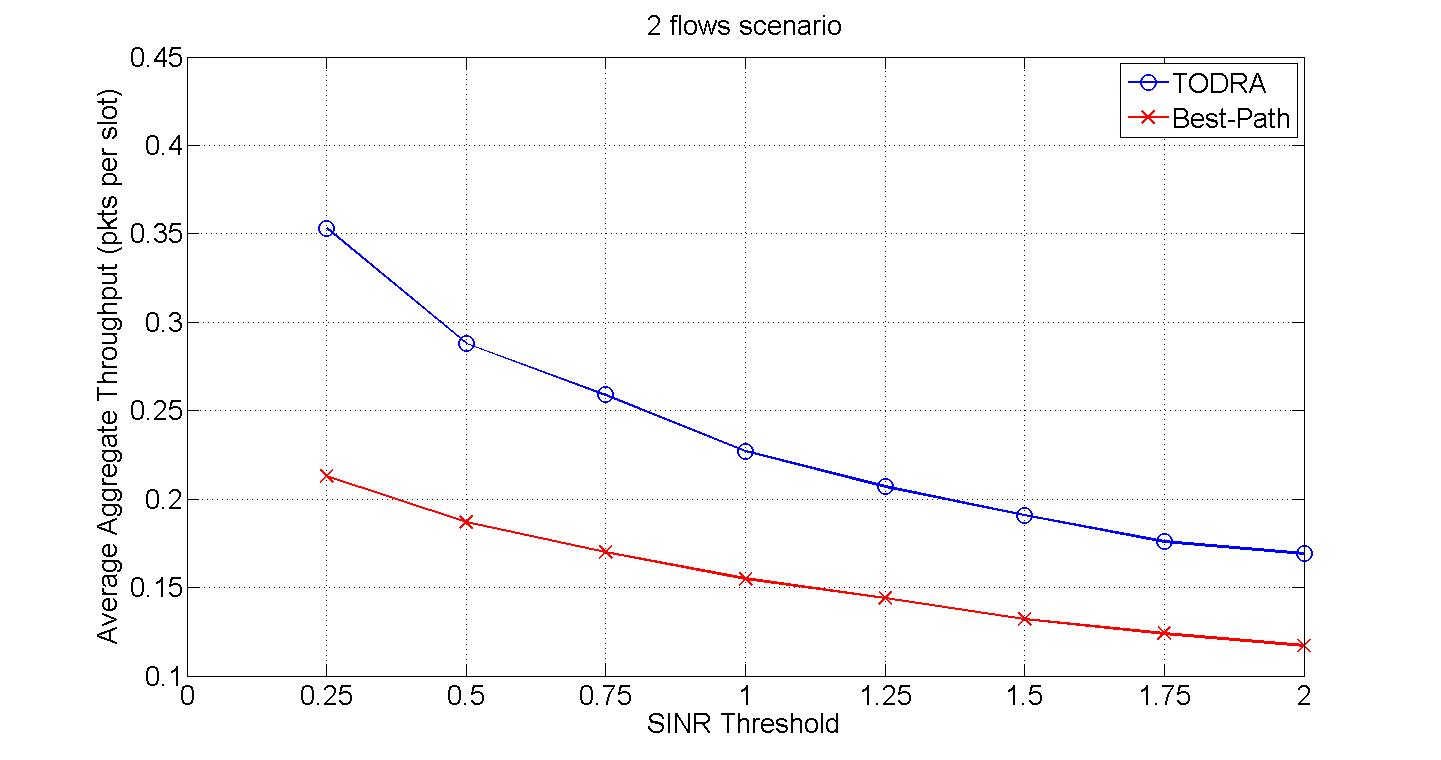}
\caption{Average Aggregate Throughput: TODRA vs Best-Path - Two flows scenario.}
\label{fig:fig_3}
\end{figure}
\begin{figure}
\centering
\includegraphics[width=9.2cm, height=6.2cm]{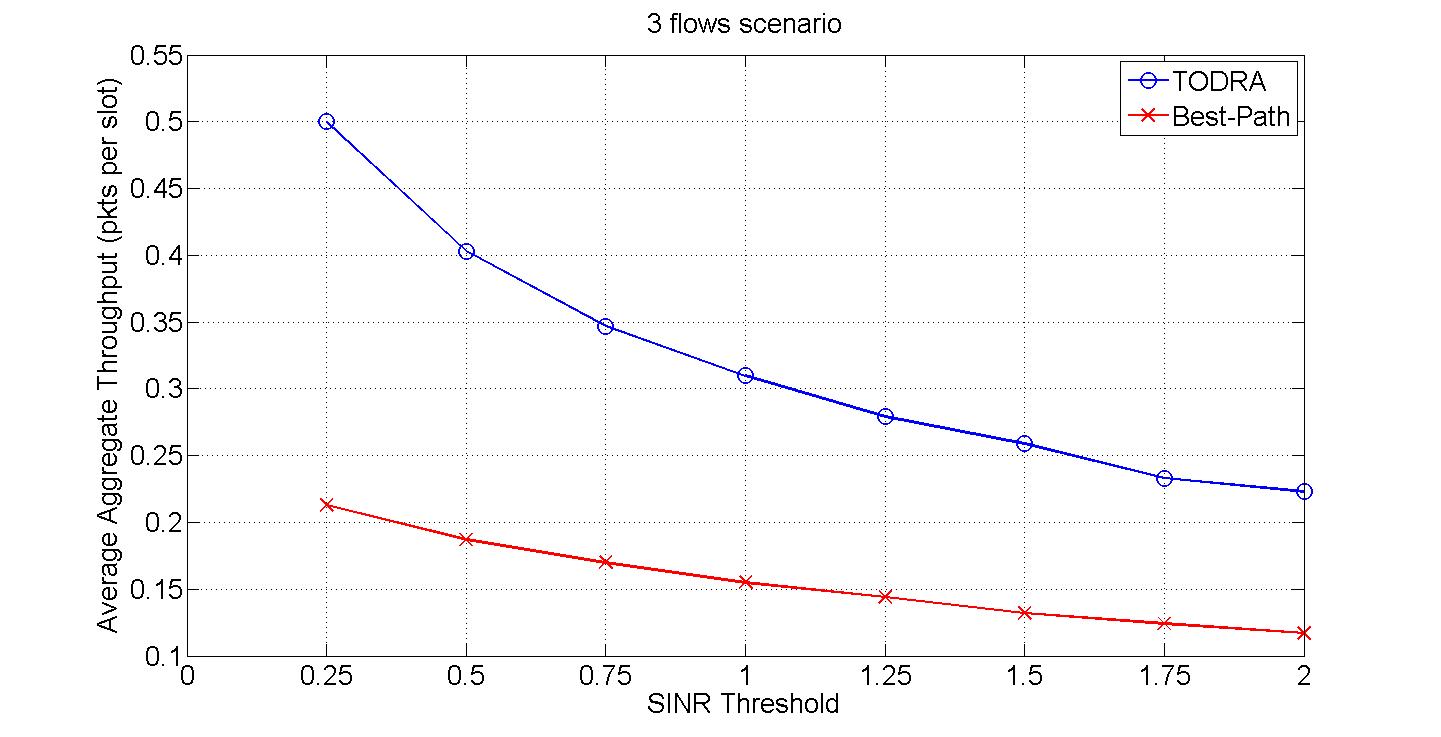}
\caption{Average Aggregate Throughput: TODRA vs Best-Path - Three flows scenario.}
\label{fig:fig_4}
\end{figure}

The second part of the evaluation process consists of comparing the suggested flow rate allocation scheme with  the Best-Path scheme.
For the topology presented in Fig. \ref{fig:grid_topology}, the path exhibiting the highest end-to-end success probability is $r_{1}$: 3-7-11-15.
Best-Path identifies the optimal data rate for the flow assigned to the path selected by solving a single path version of optimization problem P2.
Fig.~\ref{fig:fig_3}, and~\ref{fig:fig_4} compare the average aggregate throughput achieved by our scheme and Best-Path for the two and three flows scenarios and SINR threshold values $\lbrace 0.25, 0.5,...2.0 \rbrace$.
In both traffic scenarios, our scheme achieves significantly higher AAT than Best-Path.
The improvement in terms of average aggregate throughput is more profound in the scenario where three flows are employed to forward traffic to destination.
In the two flows scenario, our scheme achieves 49.1\% higher AAT on average than Best-Path concerning all SINR values.
The corresponding value for the three flows scenario is 102.8\%.
It is also interesting to observe the data rates assigned to each flow.
In the three flows scenario, flow $f_{2}$ is the one that experiences the most significant inter-path (or inter-flow) interference due to transmissions from flows $f_{1}$ and $f_{3}$.
Inter-path interference between flows $f_{1}$ and $f_{3}$ is mitigated by the topological separation of the paths utilized.
Using an SINR threshold value equal to 0.5 for example, flows $f_{1}$ and $f_{3}$ are both assigned a data rate of 0.496 packets per slot while $f_{2}$ a rate of 0.222 packets per slot.
This shows that the proposed scheme \textit{prefers} paths with lower correlation in terms of interference and tends to \textit{avoid} sub-optimal paths.
Apart from that, Fig. \ref{fig:fig_3}, and \ref{fig:fig_4} also show that even for SINR threshold values larger than one, where multi-packet reception is less probable, the suggested flow rate allocation scheme, employing multiple paths to the destination achieves significantly higher performance than Best-Path which utilizes a single path.
\section{Conclusions}
\label{sec:conclusions}

This study explores the issue of aggregate throughput optimal data rate allocation for flows forwarded over multiple disjoint interfering paths for wireless multi-hop random access networks with multi-packet reception capabilities.
We propose a distributed scheme that formulates flow rate allocation as an optimization problem.
The key feature of the suggested scheme is that it maximizes the average aggregate throughput of all flows while also providing bounded delay guarantees.
We also propose a simple model for the average aggregate throughput achieved by all flows that accounts for both intra- and inter-path interference.
Employing a toy topology we present the conditions under which the corresponding optimization problem is non-convex.
The proposed scheme is evaluated using NS-2 simulations.
Our preliminary results reveal that the suggested scheme accurately estimates average aggregate throughput despite the simplifying assumptions adopted by our analysis.
Moreover, it achieves significantly higher throughput than Best-Path allocation scheme for both traffic scenarios explored, even for SINR threshold values larger than one.

Our future steps consists of exploring more complex topologies with specific traffic patterns.
We also plan to consider multiple transmission rates and relax the assumption of fixed transmission probability for relay nodes.
In the present study we treat interference as noise and we reside in exhaustive
enumeration of the interference set of a link's receiver. In future steps we aim at adopting more sophisticated approaches
for interference handling, such as, successive interference cancelation and joint decoding \cite{b:Tse}.
%\vspace{-0.1in}
\bibliographystyle{unsrt}
%\end{thebibliography}
\bibliography{bibliography}

\begin{thebibliography}{10}

\bibitem{6133896}
Long Le.
\newblock Multipath routing design for wireless mesh networks.
\newblock In {\em Global Telecommunications Conference (GLOBECOM 2011), 2011
  IEEE}, pages 1 --6, dec. 2011.

\bibitem{mp_route_wim}
J.W. Tsai and T.~Moors.
\newblock Interference-aware multipath selection for reliable routing in
  wireless mesh networks.
\newblock In {\em Mobile Adhoc and Sensor Systems, 2007. MASS 2007. IEEE
  Internatonal Conference on}, pages 1 --6, oct. 2007.

\bibitem{mp_route_mpdsr}
R.~Leung, Jilei Liu, E.~Poon, A.-L.C. Chan, and Baochun Li.
\newblock Mp-dsr: a qos-aware multi-path dynamic source routing protocol for
  wireless ad-hoc networks.
\newblock In {\em Local Computer Networks, 2001. Proceedings. LCN 2001. 26th
  Annual IEEE Conference on}, pages 132 --141, 2001.

\bibitem{5198989}
Zijian Wang, E.~Bulut, and B.K. Szymanski.
\newblock Energy efficient collision aware multipath routing for wireless
  sensor networks.
\newblock In {\em Communications, 2009. ICC '09. IEEE International Conference
  on}, pages 1 --5, june 2009.

\bibitem{ref_mhrp}
Muhammad~Shoaib Siddiqui, Syed~Obaid Amin, Jin~Ho Kim, and Choong~Seon Hong.
\newblock Mhrp: A secure multi-path hybrid routing protocol for wireless mesh
  network.
\newblock In {\em Military Communications Conference, 2007. MILCOM 2007. IEEE},
  pages 1 --7, oct. 2007.

\bibitem{5425265}
Jiming Chen, Shibo He, Youxian Sun, P.~Thulasiraman, and Xuemin Shen.
\newblock Optimal flow control for utility-lifetime tradeoff in wireless sensor
  networks.
\newblock In {\em Global Telecommunications Conference, 2009. GLOBECOM 2009.
  IEEE}, pages 1 --6, 30 2009-dec. 4 2009.

\bibitem{flow_alloc_2008}
Kun-Da Wu and Wanjiun Liao.
\newblock Flow allocation in multi-hop wireless networks: A cross-layer
  approach.
\newblock {\em Trans. Wireless. Comm.}, 7(1):269--276, January 2008.

\bibitem{4509706}
U.~Akyol, M.~Andrews, P.~Gupta, J.~Hobby, I.~Saniee, and A.~Stolyar.
\newblock Joint scheduling and congestion control in mobile ad-hoc networks.
\newblock In {\em INFOCOM 2008. The 27th Conference on Computer Communications.
  IEEE}, pages 619 --627, april 2008.

\bibitem{kelly98ratecontrol}
F.~Kelly, A.~Maulloo, and D.~Tan.
\newblock Rate control in communication networks: shadow prices, proportional
  fairness and stability.
\newblock In {\em Journal of the Operational Research Society}, volume~49,
  1998.

\bibitem{4712692}
Ping Wang, Hai Jiang, Weihua Zhuang, and H.V. Poor.
\newblock Redefinition of max-min fairness in multi-hop wireless networks.
\newblock {\em Wireless Communications, IEEE Transactions on}, 7(12):4786
  --4791, december 2008.

\bibitem{conf/icc/QiuBX12}
Fan Qiu, Jia Bai, and Yuan Xue.
\newblock Towards optimal rate allocation in multi-hop wireless networks with
  delay constraints: A double-price approach.
\newblock In {\em ICC}, pages 5280--5285. IEEE, 2012.

\bibitem{5089987}
Qinghai Gao, Junshan Zhang, and S.~Hanly.
\newblock Cross-layer rate control in wireless networks with lossy links:
  leaky-pipe flow, effective network utility maximization and hop-by-hop
  algorithms.
\newblock {\em Wireless Communications, IEEE Transactions on}, 8(6):3068
  --3076, june 2009.

\bibitem{horizon_2008}
Bo\v{z}idar Radunovi\'{c}, Christos Gkantsidis, Dinan Gunawardena, and Peter
  Key.
\newblock Horizon: balancing tcp over multiple paths in wireless mesh network.
\newblock In {\em Proceedings of the 14th ACM international conference on
  Mobile computing and networking}, MobiCom '08, pages 247--258, New York, NY,
  USA, 2008. ACM.

\bibitem{Verdu:1998:MD:521411}
Sergio Verdu.
\newblock {\em Multiuser Detection}.
\newblock Cambridge University Press, New York, NY, USA, 1st edition, 1998.

\bibitem{b:Tse}
David Tse and Pramod Viswanath.
\newblock {\em Fundamentals of wireless communication}.
\newblock Cambridge University Press, New York, NY, USA, 2005.

\bibitem{b:Nguyen}
G.D. Nguyen, S.~Kompella, J.E. Wieselthier, and A.~Ephremides.
\newblock Optimization of transmission schedules in capture-based wireless
  networks.
\newblock In {\em Military Communications Conference, 2008. MILCOM 2008. IEEE},
  pages 1 --7, 2008.

\bibitem{ref:ns2}
{The Network Simulator NS-2}.
\newblock \url{http://www.isi.edu/nsnam/ns/}.

\bibitem{ref:dei80211mr}
SIGNET research group.
\newblock dei80211mr: a new 802.11 implementation for ns-2, 2008.
\newblock Available at http://www.dei.unipd.it/wdyn/?IDsezione=5090. [Último
  acesso: 18/01/2010].

\bibitem{ref:um-olsr}
Francisco~J. Ros.
\newblock Um-olsr.
\newblock \url{http://masimum.dif.um.es/?Software:UM-OLSR}.

\end{thebibliography}

\end{document}